# Tunable graphene phononic crystal

*Jan N. Kirchhof[1][*], Kristina Weinel[1,2], Sebastian Heeg[1], Victor Deinhart[2,3], Sviatoslav Kovalchuk[1], Katja Höflich[2,3] and Kirill I. Bolotin[1][*]*

[1]Department of Physics, Freie Universität Berlin, Arnimallee 14, 14195 Berlin, Germany.

[2]Ferdinand-Braun-Institut gGmbH Leibniz-Institut für Höchstfrequenztechnik, Gustav-Kirchhoff-Str. 4, 12489 Berlin, Germany.

[3]Helmholtz-Zentrum Berlin für Materialien und Energie, Hahn-Meitner-Platz 1, 14109 Berlin, Germany.

*jan.kirchhof@fu-berlin.de   *kirill.bolotin@fu-berlin.de

# Abstract

In the field of phononics, periodic patterning controls vibrations and thereby the flow of heat and sound in matter. Bandgaps arising in such phononic crystals (PnCs) realize low-dissipation vibrational modes and enable applications towards mechanical qubits, efficient waveguides, and state-of-the-art sensing. Here, we combine phononics and two-dimensional materials and explore tuning of PnCs via applied mechanical pressure. To this end, we fabricate the thinnest possible PnC from monolayer graphene and simulate its vibrational properties. We find a bandgap in the MHz regime, within which we localize a defect mode with a small effective mass of $0.72 \text{ ag} = 0.002 \, m_{\text{physical}}$. We exploit graphene's flexibility and simulate mechanical tuning of a finite size PnC. Under electrostatic pressure up to 30 kPa, we observe an upshift in frequency of the entire phononic system by ~350%. At the same time, the defect mode stays within the bandgap and remains localized, suggesting a high-quality, dynamically tunable mechanical system.

# Introduction

A phononic crystal (PnC) is an artificially manufactured structure with a periodic variation of material properties e.g. stiffness, mass, or stress.[1] This periodic perturbation creates a meta-crystallographic order in the system leading to a vibrational band structure hosting acoustic Bloch waves, in analogy to the electronic band structure in solids.[1] Designing the lattice parameters of the meta-structure allows to directly manipulate phonons at various length scales.[2–4] This can be used to guide[5–7] and focus phonons,[8,9] or to open a vibrational bandgap.[1,10–12]

Phononic bandgaps in periodic structures suppress radiation losses and allow for highly localized modes (of frequency $f$) on artificial irregularities.[13,14] The quality factors ($Q = \frac{f}{\Delta f}$) of these so-called defect modes are especially high.[15,16] In particular, resonances with Q > 8 x 10⁸ have been observed at room temperature in silicon nitride (SiN) PnCs.[15–17] In these devices, the quality factor exceeds the empirical $Q \sim m^{1/3}$ rule[17–19] and the vibrational periods overcome the thermal decoherence time limit: $\tau = hQ/k_B T$.[15,17] This, in turn, enables the study of quantum effects in resonators of macroscopic size – all at room temperature.[20,21]

Frequency tunability in PnCs could add an unprecedented knob to control a broad range of phononic application and thereby provides access to new regimes of guiding, filtering, and focusing phonons.[22–33] It would furthermore allow to resonantly couple to an external optical or mechanical excitation and thus realize sensing applications with mechanical qubits and studies on quantum entanglement.[34] Yet, the mechanical resonances in PnCs are determined by material constants and the crystal geometry.[22,23,26–28] In principle, the mode frequencies can be controlled by changing the temperature[29,30] or by an external magnetic field.[31,32] This, however, only provides limited tunability and necessitates heating the system or inclusion of magnetic materials. While SiN, as well as other conventional low-loss materials, is very stiff and allows only limited mechanical tunability,[24,33] strain has been used to adjust the frequency response of elastic polydimethylsiloxane (PDMS).[25] Unfortunately, low crystalline quality of that material led to limited tunability and very small Qs for mechanical modes.

Recently, PnCs made from two-dimensional (2D) materials have been considered.[35–37] Such materials feature intrinsically low mass, high fundamental frequency, and easily accessible displacement non-linearity. Most importantly, their high tensile strength and monolayer character allows to mechanically strain them up to ten percent.[38] That invites consideration of mechanically controllable 2D-material based PnCs. Specifically, we expect the entire acoustic band structure of such a PnC to be highly tunable by applying mechanical pressure. Nevertheless, tunability of 2D phononic systems as well as localized defect modes in them have not been studied yet.

Here, we investigate mechanical tunability in a realistic graphene PnC. We fabricate a suspended micron-sized monolayer graphene PnC via focused helium ion beam milling (FIB) and characterize it spectroscopically. We then use experimentally established parameters to calculate the phononic band structure of the resulting PnC. We find a phononic bandgap from 48.8 to 56.5 MHz inside of which we localize a defect mode with an effective mass of 0.72 ag. Finally, we computationally investigate the mechanical tunability of the PnC under pressure induced by a local electrostatic gate.[39,40] The applied pressure smears out the phononic bandgap as the out-of-plane displacement breaks the symmetry and causes perturbations of the artificial lattice, yet the mode shape of the defect mode remains highly localized. Overall, we can tune the resonance frequency of the defect mode by more than 350% and access new regimes of strain engineering.

## Results

**Designing a tunable phononic crystal**

Our device design of a tunable, two-dimensional PnC consists of the following key elements. First, the PnC material must be freestanding to allow out-of-plane displacement. Second, it is necessary to use an electrically conductive material. In that case, an electrostatic gate electrode can be used to apply pressure and to induce tension as the membrane is pulled towards the gate. Third, the material needs to be flexible to allow large mechanical tunability with small pressures. Monolayer graphene with its high carrier mobility >200.000 cm²/Vs[41] and large breaking strength >10%[38] perfectly fulfils these requirements. By using large area CVD graphene we can fabricate many devices on a single chip. Finally, the device needs to host a large enough number of unit cells with sufficient periodicity to form a well-defined PnC. While this task is simple in thick SiN, it is much more challenging for fragile freestanding monolayer graphene. To overcome this, we choose a much smaller unit cell compared to typical SiN-PnCs (~100 µm size) and use helium FIB-milling to pattern the PnC.[42] This direct lithography allows to pattern graphene down to 10 nm features,[43] whilst causing little damage.[44,45] A patterned prototype monolayer graphene PnC is shown in Figure 1A. It consists of a honeycomb lattice of holes (lattice constant $a = 350$ nm, hole diameter $d = 105$ nm) around a central region. Within its 10 µm diameter the two-dimensional PnC contains more than 30 unit cells. The honeycomb lattice inspired by Tsaturayn et al.[15] exhibits a robust bandgap[12,15,46] whilst retaining a relatively large fraction of material to ensure a stable device. Additional PnC with various patterning sizes are shown in Figure S1–S3.

Next, we map the tension within the produced structures using Raman spectroscopy. We expect tension hot spots in the thin ribbons and relaxation in the centers of the hexagons.[47] Such tension redistribution should affect the vibrational properties of our PnC. To this end, we fabricate another prototype device (Figure 1B) with lattice constant $a = 2$ µm and spatial features comparable to the size of a focused laser spot. The intensity map of the 2D-Raman mode of graphene for this device is shown in Figure 1C. The intensity of the 2D-mode corresponds to the amount of material while its spectral position depends on the tension in the material.[48,49] In the pizza-like image one can clearly see the removed material from the drop in intensity and identify the honeycomb lattice. In Figure 1D, we compare the spectral position of the Raman 2D-mode for a graphene PnC (blue) along the dashed line shown in Figure 1C to an unpatterend graphene membrane (red). The quasi-periodic variations in the PnC device, that are absent in the unpatterned reference, correspond to the redistributed tension. In Figure 1E we compare the extracted relative tension (blue) to a simulation (yellow) and find the expected signatures of tension redistribution – higher tension between the holes and lower tension in the middle of the hexagons (details in Supporting Information).

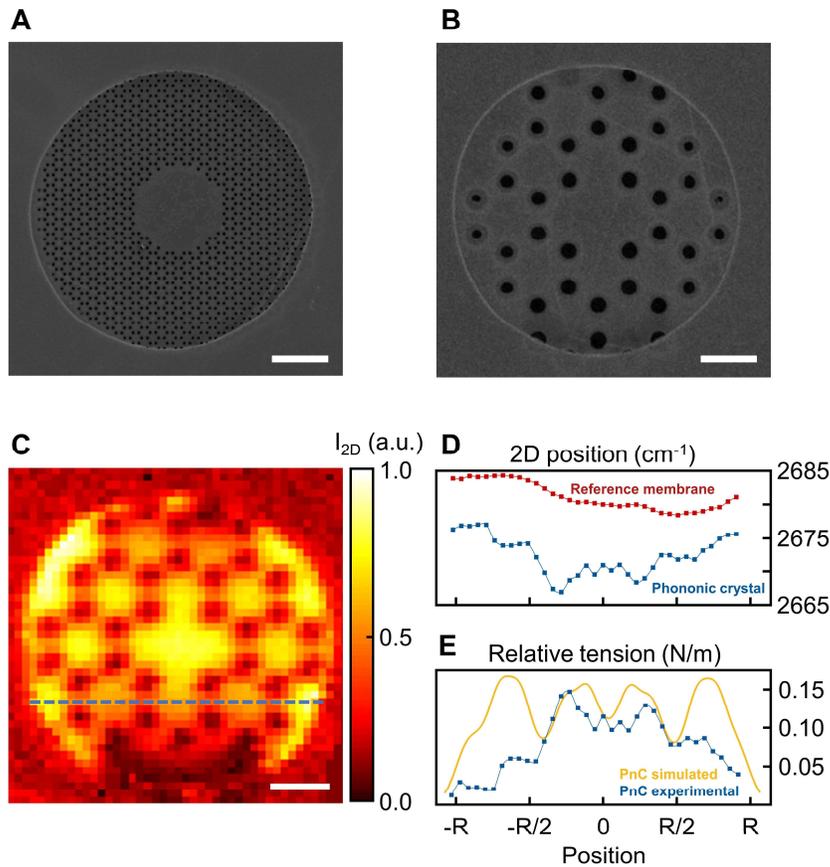

**Figure 1. Graphene phononic crystals and tension redistribution.** (**A,B**) Helium ion micrographs of prototype monolayer graphene phononic crystal devices with lattice constants 350 nm and 2 µm, respectively. Scale bar length is 2 µm. The phononic pattern, a honeycomb lattice of holes with a defect in its center, allows us to localize a vibrational defect mode. The ring-like features around the holes in (**B**) are due to incomplete removal of graphene most probably caused by contamination (details in Supporting Information). (**C**) Intensity map of the Raman-active 2D mode of graphene for the device shown in (**B**). The periodic pattern is clearly visible. (**D**) Raman 2D-mode position along a line cut (dashed line in (**C**)) for a PnC (blue) and reference membrane

(red). The PnC shows a periodic variations of much larger amplitude compared to the fluctuation in the reference sample. (**E**) Comparison of the relative tension extracted from Raman measurements (blue) to the simulated tension distribution (yellow) confirming the redistribution of tension upon pattering. The simulation includes spatial broadening due to the finite size of the laser spot.

**Phononic crystal simulations**

Having experimentally established the feasibility of a suspended graphene PnC, we use our findings to simulate its phononic properties in two independent approaches. First, we calculate the phononic band structure for an infinitely repeated unit cell ("infinite model"). This model is well-accepted and fast.[15–17] However, due to the size limits of suspended graphene, our devices are smaller than typical SiN-PnCs (mm size)[15–17] and contain fewer unit cells. Furthermore, we want to apply pressure to the entire system and investigate localized modes in the bandgap. Therefore, we also simulate a more realistic system of finite size ("finite model"). For both models, we use the honeycomb lattice with feasible parameters and account for tension redistribution upon fabrication (Figure 1, D and E). We choose a lattice constant $a = 1$ μm, a filling factor of $d/a = 0.5$ (slightly larger than in Figure 1) and an initial tension of $T_0 = 0.01$ N/m, a realistic value for clean monolayer graphene.[39,50]

**Infinite model**

By applying periodic boundary conditions to the unit cell (Figure 2A) we calculate the band structure for an infinite honeycomb lattice (Figure 2B). We find a mixture of in-plane (dashed lines) and out-of-plane modes (solid lines). From the slope of the out-of-plane modes in Figure 2B, we determine the speed of sound $v_g = \frac{\partial \omega}{\partial k} = 83$ m/s. In the range from 48.8 to 56.5 MHz (red shaded area) we find a bandgap for out-of-plane modes. This quasi-bandgap (in-plane modes are still present) has a gap-to-midgap ratio of 14.6%. The in-plane modes do not couple to out-of-plane modes[51] and therefore do not hinder radiation shielding. The bandgap originates from Bragg scattering, with each hole acting as a scatterer for out-of-plane oscillations. Upon negative interference conditions, directional Bragg bandgaps open at the high symmetry points. Where these gaps overlap radiation shielding becomes possible, as wave propagation is isotropically forbidden.[1] The bandgap position depends reciprocally on $a$. With our fabrication schema we can tailor the bandgap center from 350 to 26 MHz by varying $a$ from 0.175 to 2 µm (Figure 2C, devices in Figure S2,S3). Overall, the simulations in the infinite model suggest the possibility of a large quasi-bandgap, which we will next use to control phonons.

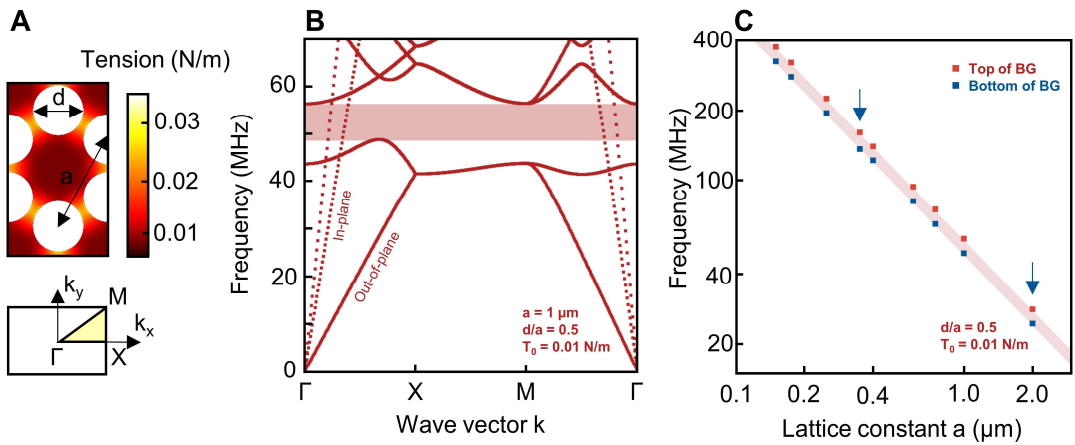

**Figure 2. Band structure calculations of an infinite graphene phononic crystal.** (**A**) Unit cell of the honeycomb lattice with redistributed tension (top) and the corresponding first Brillouin zone (bottom). (**B**) Phononic band structure for the unit cell shown in (**A**). In-plane modes are shown as dashed lines, out-of-plane modes as solid lines, and the corresponding quasi-bandgap region as red shaded area. (**C**) Top (red) and bottom (blue) of the bandgap vs. lattice constant. The blue arrows indicate the lattice constant of the devices from Figure 1.

**Finite model**

To study a realistic device of finite size under electrostatic pressure and to implement a defect into the phononic pattern, we conduct a second independent simulation ("finite model"). In this model, we consider a finite number of unit cells of the honeycomb lattice (same $a$, $d/a$, and $T_0$ as before) and employ fixed boundary conditions along the PnC's perimeter. We choose a circular device as such a geometry allows uniform suspension and minimizes edge effects. In the center of the 30.6 μm device we create a 1.9 μm hexagonal defect,[15] as sketched in Figure 3A. Freestanding graphene devices of that size have been fabricated[52] and the central defect area is large enough to measure resonances interferometrically.[53,54] Next, we simulate the first 1500 eigenfrequencies and the corresponding spatial mode shape. In Figure 3B, we plot the frequencies $f$ vs. mode number $N$ for the PnC (blue) and compare it to an unpatterned graphene membrane as reference (green). The graph for the PnC shows signs of a bandgap, as we observe an initial flattening of the curve followed by a sudden increase. This region of reduced mode density coincides exactly with the bandgap from our infinite model (blue area) and stands in contrast to the unpatterned membrane for which the frequencies gradually increase with mode number. The second indication of the bandgap is evident when we examine the effective mass of the modes:

$$m_{\text{eff}} = \rho_{2D} \iint \frac{z^2}{z_{\text{max}}^2} dxdy,$$

where $\rho_{2D}$ is the areal density of graphene and $z$ ($z_{\text{max}}$) is the (maximum) vibration amplitude in z-direction. For the fundamental mode we obtain $m_{\text{eff}} = 80.9$ ag $= 0.252\ m_{\text{physical}}$, which roughly matches the literature value for the mode shape of a uniform, circular membrane of $m_{\text{eff}} = 0.269\ m_{\text{physical}}$.[55] We observe a pronounced drop of $m_{\text{eff}}$ in the bandgap region (Figure 3C). This observation is consistent with localized modes inside the bandgap, which typically show a small average displacement resulting in a reduced effective mass.[17]

Finally, we directly extract the band structure from the results of the finite model and compare it to that of the infinite model. To accomplish this, we analyze the mode shape of each resonance following Ref [56]. Specifically, we take the spatial FFT of each mode shape to find its representation in reciprocal space and to assign a wave vector $k$ to each mode. In Figure 3 E-H, we show real space (top) and reciprocal space (bottom) plots of exemplary modes. Mode I (20.2 MHz – Figure 3E) is below the bandgap and resembles a higher order Bessel mode in real space, which transforms to a near-uniform circle in momentum space. A higher frequency mode IV (60.7 MHZ – Figure 3H) is situated above the bandgap. For this mode, we observe zone folding as the mode reaches out beyond the 1.BZ (dashed line). Analyzing all 1500 modes lets us restore the dispersion relation beyond the 1.BZ (Figure 3D, blue), which almost perfectly matches the band structure from the infinite model (red). From our observations of reduced mode density (Figure 3B), drop in effective mass (Figure 3C), and mode shape-analysis (Figure 3D), we confirm the presence of a bandgap for out-of-plane modes in a realistic system of finite size.

Next, we examine the modes located within the bandgap and identify the defect mode. In Figure 3G, we show a typical bandgap mode in real (top) and k-space (bottom). As most modes in the bandgap, this mode is localized at the edges of the PnC in the real space. However, one mode at frequency 49.9 MHz is localized at the central defect (Figure 3F) and surrounded by the phononic pattern. We therefore identify it as our defect mode of special interest. The $m_{\text{eff}}$ of the mode is 0.724 ag, which is more than a factor 100 smaller than the fundamental mode of the system and orders of magnitude lower than for any reported SiN defect mode.[15–17] Overall, our model confirms the vibrational bandgap for a system of finite size and a localized defect mode within that bandgap.

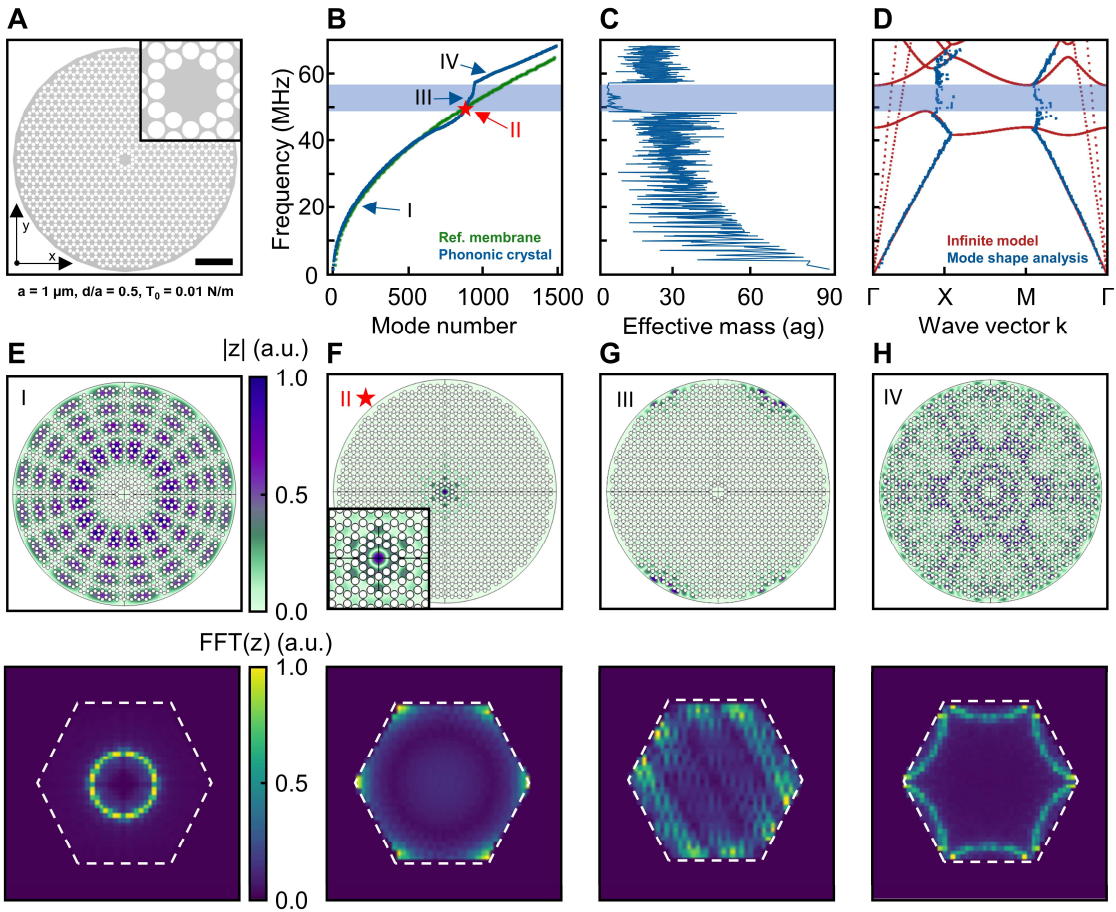

**Figure 3. Finite size model of a graphene phononic crystal.** (**A**) Device geometry for the finite system simulations (scale bar is 5 μm). A central "defect" region is designed to localize one vibrational mode and decouple it from its environment. (**B**) The first 1500 simulated eigenfrequencies vs. mode number for a PnC device (blue) and a circular membrane without patterning (green). The bandgap region from the infinite model is shown in blue. (**C**) Effective mass for each mode. The modes within the bandgap (blue) show a more than 100-fold decrease in effective mass compared to the fundamental mode. (**D**) Band structure calculated from the finite model via mode-shape analysis (blue) along with the band structure from the infinite model (red). The low energy acoustic branches fit well, and the bandgap regions coincides with the simulated results from the infinite model (red). (**E-H**) Exemplary mode shapes in real (top) and reciprocal space (bottom) for: (**E**) a mode below the bandgap (I), (**F**) the defect mode (II), (**G**) another highly localized mode in the bandgap (III) and (**H**) a mode above the gap (IV).

**Phononic crystal tuning**

We now show the key advantage of our graphene PnC – dynamic and rapid frequency tuning of the bandgap and of the defect mode. To demonstrate this, we model our graphene PnC under pressure, which is applied by an electrostatic gate. The pressure causes displacement of the suspended membrane and increases the in-plane tension. We initially approximate this effect in first order in our infinite model by neglecting out-of-plane displacement and simply increasing the in-plane tension. In Figure 4A, we plot the band structure for $T_0 = 0.010$ and $0.012$ N/m. We observe a frequency increase of the out-of-plane modes and thus an upshift of the quasi-bandgap by 10%. The speed of sound $v_g$ rises from 83 to 830 m/s in the range of tension from 0.01 to 1 N/m (Figure 4B). The system (finite and infinite) behaves like a thin membrane under tension and the frequencies of the PnC scale directly with tension: $f \propto \sqrt{\frac{T_0}{\rho_{2D}}}$.[55] This scaling makes our system highly sensitive to tension and in combination with graphene's mechanical flexibility allows for broad frequency tuning.

Having demonstrated the overall tunability of our system, we now simulate the effect of electrostatic pressure on the phononic system and the defect mode in a realistic device. To do so, we switch to the finite model and apply pressure in negative z-direction. In our simulations we stick to experimentally reported pressure values and apply a maximum of 30 kPa.[39] To investigate the influence of pressure on the bandgap, we compute the phononic density of states, $DOS = dN/df$, and plot it vs. pressure (Figure 4C). In this plot, the bandgap is distinguished by a reduced DOS. While at zero pressure the bandgap region is obvious, for higher pressures the drop becomes less pronounced (Figure 4C). We attribute this smearing to a breaking of symmetry, perturbation of the PnC as it deforms under pressure (inset Figure 4C), and rising non-uniformity in the tension distribution (Figure S6E). Nevertheless, we estimate the top and bottom of the bandgap, Figure 4D (blue). A bandgap tuning by more than 300% is evident. We verify the bandgap tuning by an independent approach based on averaging the induced tension (red markers, details in Supporting Information).

Next, we investigate tunability of the defect mode. Upon applying 30 kPa pressure to a device with an initial tension of 0.01 N/m, the resonance frequency of the defect mode upshifts from 49.9 to 217.5 MHz (black stars Figure 4D). Since the bandgap is smeared under pressure (Figure 4C), it is important to check the localization of the defect mode. Hence, we inspect a line cut through the center of the device and plot the normalized mode shape vs. pressure in Figure 4E. The shape as well as the effective mass (inset Figure 4E) of the mode remains virtually unchanged and the mode retains its localization. Summarizing, we have shown a tunable speed of sound and realized an upshift of the defect mode resonance under pressure, whilst maintaining its localization. Such a more-than-four-fold frequency increase is unprecedented and remains elusive in any other phononic systems.[22–33]

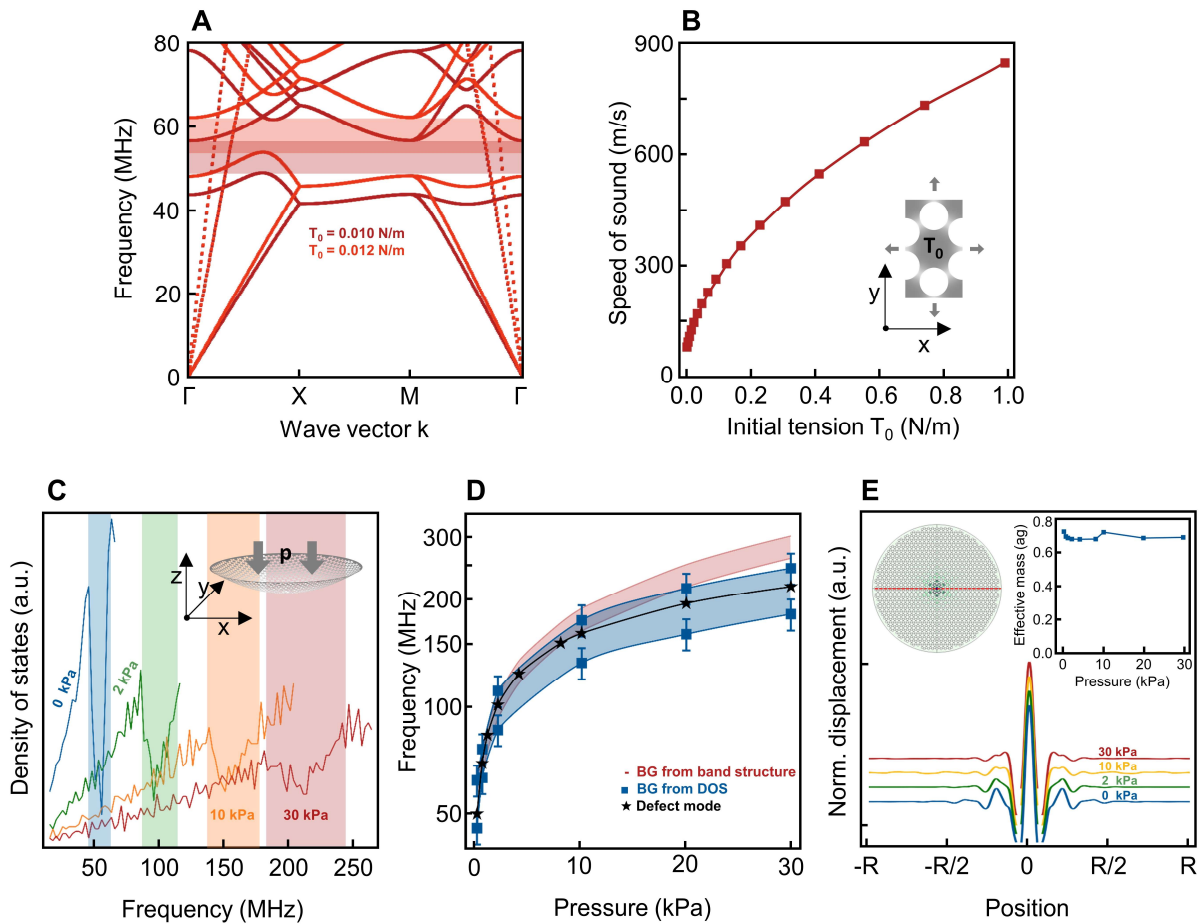

**Figure 4. Mechanically tunable graphene phononic crystal.** (**A**) Band structure for initial tension values $T_0 = 0.010$ N/m (red) and $T_0 = 0.012$ N/m (orange). The entire out-of-plane branch scales strongly with tension. The position and width of the bandgap are equally tension-dependent. (**B**) Speed of sound for the out-of-plane modes extracted from (**A**) vs. tension. (**C**) Density of states calculated from the finite model as a function of pressure applied to the suspended PnC ($T_0 = 0.010$ N/m). (**D**) Pressure dependence of resonance frequency of the central defect mode (stars), of the bandgap from infinite model (red shaded), and of the bandgap extracted from the density of states (blue squares). The defect mode remains within the bandgap even at high pressures. (**E**) Line cut for the spatial profile of the defect mode at different pressures (vertically offset for clarity). Even at large applied loads, the mode shape remains localized and the effective mass (inset) stays constant.

## Discussion

We now discuss experimental signatures of this system. The spatial features of the extended modes in our device (Figure 3, E and H) are too fine to be resolved via diffraction-limited optics. At the same time, the extent of the defect mode is in the size of microns (Figure 3F) allowing the detection of that mode via interferometric read-out (Figure S8).[53,54] This mode has a non-zero net displacement and can be directly actuated via electrostatic drive. It will be straightforward to distinguish the defect mode from other modes by its localization in the center of the device and its likely increased quality factor. Indeed, the quality factor is defined by: $Q = 2\pi\, E_{\text{stored}}/E_{\text{diss}}$, where $E_{\text{diss}}$ is the dissipated energy per oscillation including all dissipation mechanisms and $E_{\text{stored}}$ is the mode's total energy. As the mode shape shows zero displacement near the clamping points, we expect strongly suppressed bending losses and thus enhanced Qs. Additionally, the phononic shield hinders radiation losses into the substrate, which become especially important at higher frequencies.[16] While bending and radiation losses may play a secondary role among the mechanisms lowering $Q$ in graphene resonators, our experiments nevertheless should determine the contribution of these mechanisms. Finally, by applying pressure, we increase the stiffness of the resonator. This increases the energy stored in the system[17] and supposedly further enhances the quality factor. The demonstrated level of strain control in our system invites future studies on dissipation dilution via strain engineering following the work of Ghadimi et al.[17]

We also note that our results can be easily extended to the entire family of two-dimensional materials. Currently it is challenging for us to experimentally achieve sufficient uniformity in the graphene membrane in order to generate a spatially uniform bandgap and localize the defect mode. Monolayer graphene is rather sensitive to surface corrugations[39] and transferred CVD graphene is often covered by fabrication residues, so using thin exfoliated graphene multilayers could be a solution for which we expect to find experimental signatures. The increased uniformity in multilayer graphene comes along with a decreased tunability, yet we anticipate more than 100% relative tuning for up to ~35 layers (Figure S9). For our graphene PnC, we do not expect to reach $Q$s comparable to SiN. Nevertheless we estimate $m_{\text{eff}}$ of our defect mode to be at least eight orders of magnitude lower than in other 2D-SiN-PnCs.[15] This immensely increases the measurement rate of quantum states $\Gamma_{\text{meas}} \propto 1/m_{\text{eff}}$ and decreases thermomechanical noise.[15] The frequencies in our system are controlled by simply adjusting a gate voltage, and we expect the tuning to take place on time scales comparable to regular graphene resonators and therefore achieve tuning bandwidths > 15 kHz.[57]

## Conclusion

In summary, we have fabricated and simulated a tunable PnC made from monolayer graphene. For an experimentally-informed honeycomb lattice structure, we find a robust vibrational bandgap in the MHz range. The bandgap persists for a finite-size system and we use it to localize a defect mode and shield it from its surroundings. This defect mode shows a very small effective mass of 0.724 ag, orders of magnitude smaller compared to traditional PnCs. As our central result, we demonstrate a frequency upshift of the defect mode as well as the entire phononic system by more than 350% by applying an experimentally feasible pressure of 30 kPa. While the bandgap smears out due to out-of-plane displacement perturbing the lattice, the defect mode stays within the bandgap and remains highly localized. We suggest experimental signatures of the defect mode allowing its differentiation from other modes in the system. Overall, our design of a 2D-material based PnC adds a new knob to dynamically and rapidly tune frequencies in a broad range of phononic applications. Our results invite future experiments as our approach allows adjustable coupling of a PnC to external systems and may lead to better understanding of the dissipation mechanisms in graphene.

## Methods:

### Device fabrication

The pattering of the CVD grown graphene membranes was carried out in a He-Ion microscope (Orion Nanofab). Supporting Information section I. provides a detailed process description.

### Raman Spectroscopy

Raman mapping was performed on a Horiba Xplora Raman spectrometer using a 100x (NA 0.9) objective and 532 nm excitation. Spectra were acquired with a laser power of 0.5 mW and an integration time of 3s. Tension (via strain) values were derived from the 2D-mode position following standard procedures, see Supporting Information section IV.

### Simulations

For the finite element modelling we use COMSOL Multiphysics (Version 5.5) and assume the following material parameters for monolayer graphene: Young's modulus $E_{2D}$ = 1.0 TPa[38], Poisson's ratio of $\nu = 0.15$, thickness of $h = 0.335$ nm and a density of $\rho = \frac{\rho_{2D}}{h} = 2260$ kg/m³. The initial tension $T_0 = 0.01$ N/m thus corresponds to an initial strain: $\epsilon_0 = \frac{T_0}{E_{2D}} \approx 0.003\%$. For details see Supporting Information section II. + III.

## Supporting Information

Graphene pattering using He-FIB milling, details of the FEM-simulations and the mode shape analysis, detailed Raman spectroscopy analysis and simulations on experimental signatures.

This material is available free of charge via the internet at **http://pubs.acs.org**.

## Acknowledgments

**Funding:** This work was supported by ERC Starting grant no. 639739 and DFG TRR 227. V.D. and K.H. acknowledge financial support from DFG under grant no. HO 5461/3-1. The He ion beam patterning was performed in the Corelab Correlative Microscopy and Spectroscopy at Helmholtz-Zentrum Berlin.

**Author contributions:** J.N.K. conceived the idea. Suspended graphene devices were fabricated by K.W., S.K. and J.N.K. He-FIB pattering procedures were developed and carried out by K.H. and V.D. at HZB Berlin. S.H. acquired and analyzed Raman spectroscopy data. Sample design and FEM-modelling was performed by J.N.K. with participation by K.W. J.N.K. and K.I.B. co-wrote the paper with input from all authors. K.I.B. supervised the project. All authors discussed the results.

**Competing interests:** The authors declare no competing interests.

**Data and materials availability:** All data needed to evaluate the conclusions in the paper are present in the paper and/or the Supporting Information.

# Supporting Information: Tunable graphene phononic crystal

*Jan N. Kirchhof[1*], Kristina Weinel[1,2], Sebastian Heeg[1], Victor Deinhart[2,3], Sviatoslav Kovalchuk[1], Katja Höflich[2,3] and Kirill I. Bolotin[1*]*

[1]Department of Physics, Freie Universität Berlin, Arnimallee 14, 14195 Berlin, Germany.

[2]Ferdinand-Braun-Institut gGmbH Leibniz-Institut für Höchstfrequenztechnik, Gustav-Kirchhoff-Str. 4, 12489 Berlin, Germany

[3]Helmholtz-Zentrum Berlin für Materialien und Energie, Hahn-Meitner-Platz 1, 14109 Berlin, Germany.

*jan.kirchhof@fu-berlin.de     *kirill.bolotin@fu-berlin.de

### I. Sample synthesis and device pattering using He-FIB milling

Monolayer graphene was synthesized on copper by low pressure chemical vapor deposition (CVD). Upon reaching the growth temperature of 1035 °C a mixture of methane (5 sccm), hydrogen (10 sccm), and argon (5 sccm) was introduced into the CVD chamber. After 7 min growth time and a rapid cooldown, graphene was wet transferred onto a perforated SiN membrane, covered by Cr/Au (5 nm/35 nm) to electrically contact the graphene.

To pattern the suspended graphene, we used a beam of focused helium ions in the Orion Nanofab microscope. The local material removal upon ion beam impact is a complex interplay between physical sputtering (ions kick out surface atoms), the redeposition of the sputtered surface atoms in the close vicinity, chemical reactions (like polymerization of organic residues by the generated secondary electrons) as well as the introduction of heat and amorphization.[1] In case of two-dimensional material only physical sputtering contributes. The holes were patterned with a dwell time of 1.5 ms and a pixel spacing of 1 nm at a beam current of 4-5 pA (device settings: $2\times10^{-6}$ Torr He , $U_{acc}$ = 30 kV, UBIV = 34 kV, aperture 2 µm). The holes on the outside of device were cut first, following a spiralling milling strategy to the centre of the suspended area (Figure S1B). Here each single hole is milled in an opposite spiral order – starting at the centre of the hole (Figure S1C). If the graphene layer is completely intact and free of contamination, the process is highly reproducible (see Figure S2). In Figure S3A we show fabricated phononic crystal devices with varying lattice constant $a$: 0.175…2 µm. While the patterning allows for highly flexible variation of geometrical parameters, like lattice constant and hole diameter, the process of He-ion induced physical sputtering is highly sensitive to surface contamination. In Figure S3B this effect becomes visible by the bright regions around the spot. Here, the secondary electrons induced polymerization of the organic residues and therefore material build-up (deposition). The amount of contamination increases the minimum dose to achieve a full cut and may even dominate over physical sputtering as shown here for the smallest spots. As the coverage of organic residues may locally vary also the required minimum dose for complete graphene removal can vary locally. This can be seen in Figure 1B of the main manuscript. Holes with sizes down to 5 nm can be fabricated by He ion beam milling in relatively uniform and uncontaminated monolayer graphene (Figure S3B).

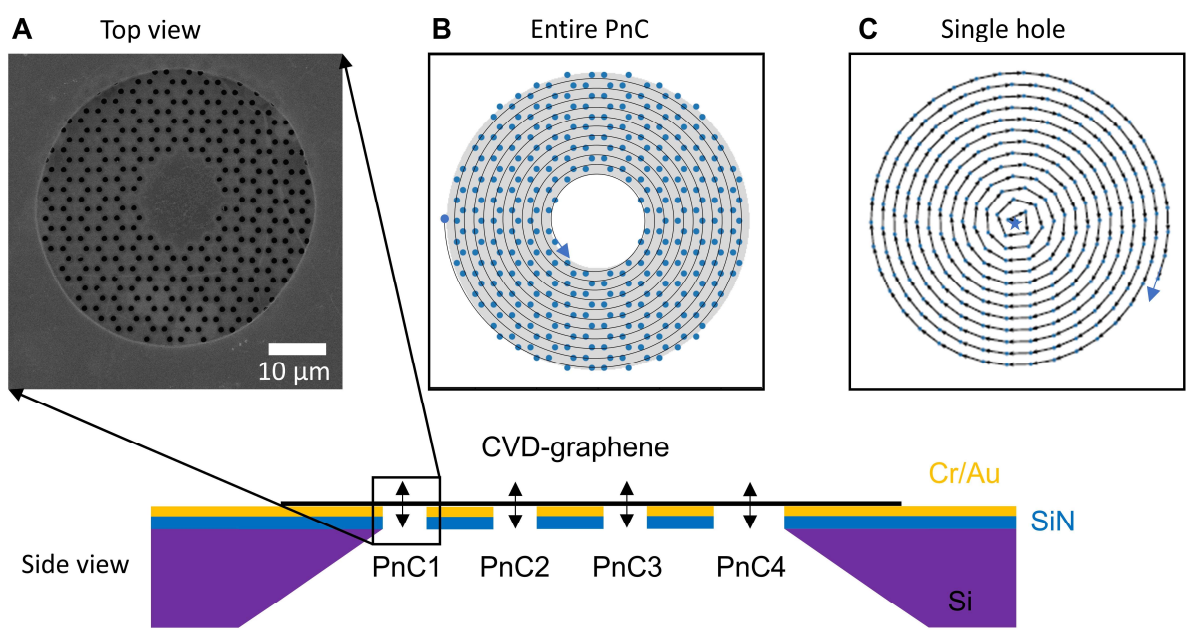

**Figure S1. Milling strategy for PnC-device patterning.** (**A**) Device with lattice constant $a = 700$ nm. By transferring CVD graphene onto pre-pattered substrates with many through holes, many PnC can be milled on a single chip. (**B**) Corresponding design file including the pattering order – starting at the outside and spiralling towards centre of the device. (**C**) Patterning of a single hole – starting in the centre moving to the outside.

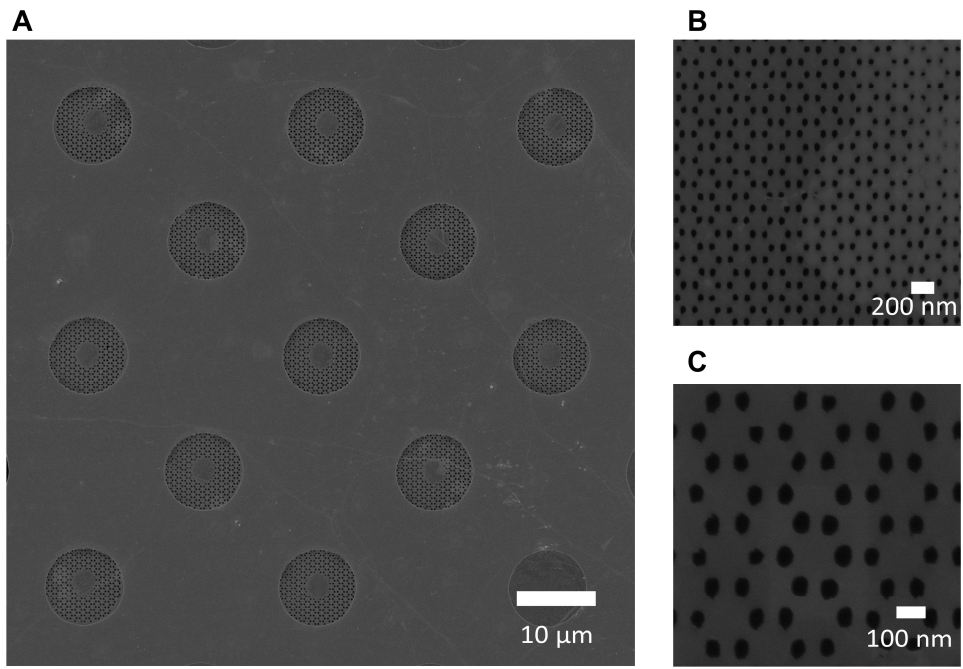

**Figure S2. Reproducible PnC patterning.** (**A**) Repeated pattering of a device with lattice constant a = 700nm. For uniform graphene the process is highly reproducible. (**B**,**C**) Zoom-in on the honeycomb lattice with a = 175 nm. The milling process is less efficient on add-layer regions – visible on the right half in (**B**).

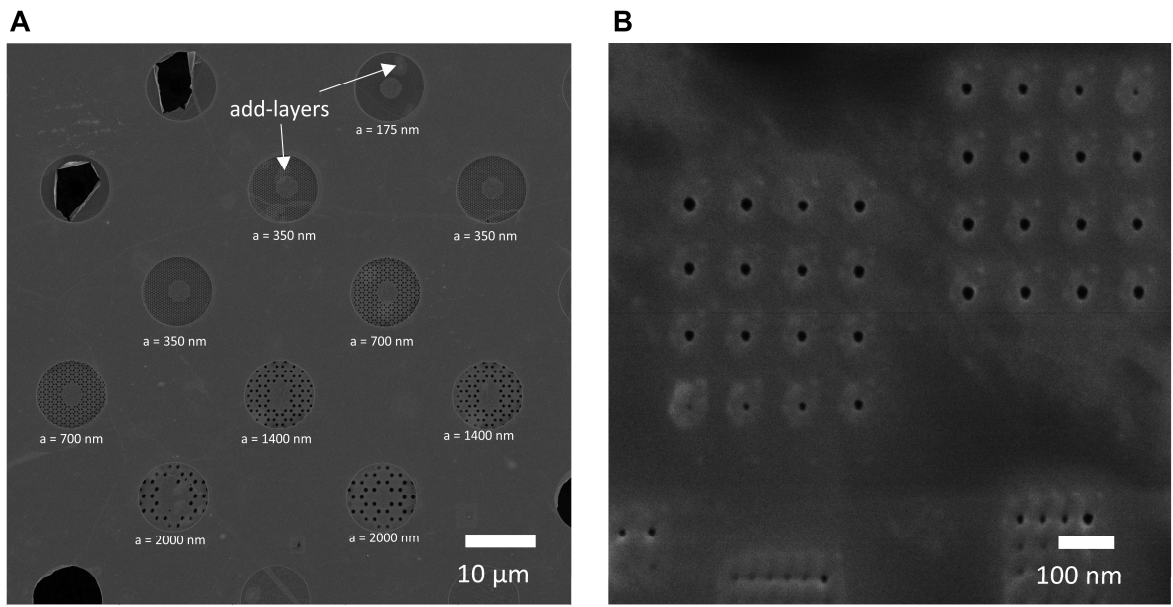

**Figure S3. Scalable PnC patterning.** (**A**) PnC devices of varying lattice constant a = 175 nm - 2 μm. Occasional add-layer regions from CVD growth are marked. (**B**) Dose tests on uniform and clean graphene, show that pores with sizes well below 5 nm can be fabricated by He ion beam milling.

## II. Finite element method simulations

We use the solid mechanics module of Comsol Multiphysics (Version 5.5) to carry out the FEM-simulations presented in the main paper.

The **infinite model** for the band structure calculations consist of two studies within one model. We use a large square containing many unit cells of the phononic pattern and implement uniform tension in the solid (Figure S4). In a stationary study with fixed boundary conditions at the edges, we simulate the tension redistribution, which occurs upon cutting holes into a system under tension.

We then add a second study (eigenfrequency domain) within in the same model to simulate the resonances und thus plot the band structure. To accurately depict the tension distribution, we crop the central unit cell of the large square from the first study and component-wise transfer the tension distribution to the second study (Fig S4). To obtain the band structure, we apply periodic boundary conditions (Floquet) to the edges of the unit cell and parameterize $k_x$ and $k_y$ (in an auxiliary sweep) along the high symmetry lines in the first Brillion zone and calculate the first 6-10 eigenfrequencies for every value of $k$. By plotting the frequencies $f$ vs. $k$, we get the dispersion relation for the geometry of interest. We use a swept mesh as we simulate a very thin system. Furthermore, we apply copy operators within in the unit cell when building the mesh to completely capture the symmetry of the system (Fig S5B). In general, the size of the bandgap depends on the filling factor $d/a$. Choosing $d/a$~0.5 (slightly larger than for Figure 1 in the main paper) results in a reasonably sized bandgap, whilst leaving behind enough material to reproducibly fabricate devices. In Figure S5, we provide a detailed study of bandgap width vs. $d/a$. Taking into account for tension redistribution overall reduces the size of the bandgap. For the second estimate of the bandgap tuning with applied pressure (main text,

Figure 4D) we extract the average tension in the finite model at each pressure value and feed the average values as input into our infinite model.

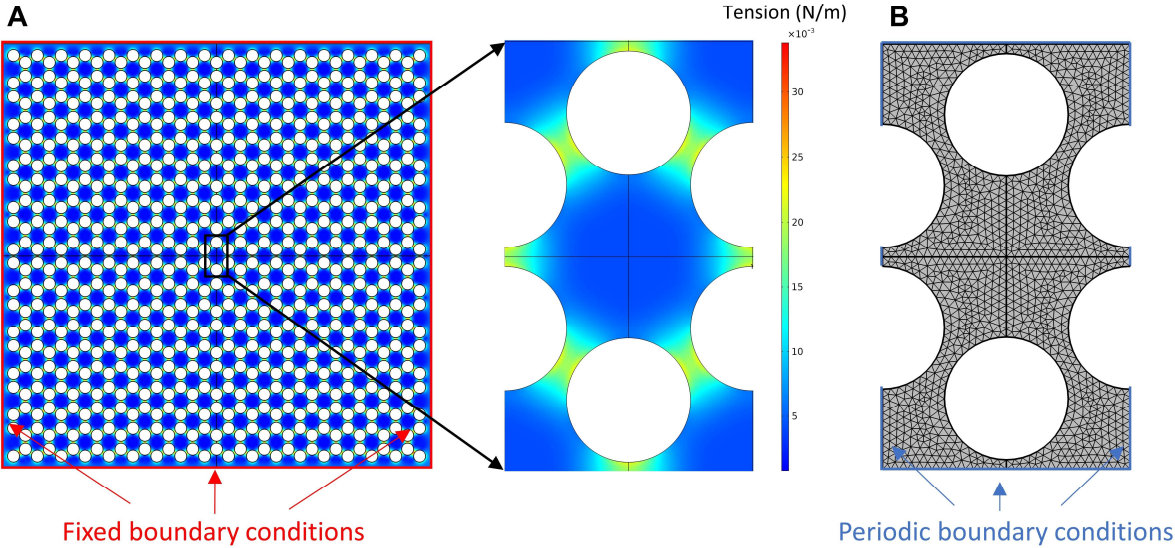

**Figure S4. Infinite model PnC simulations.** (**A**) A large membrane is needed to correctly calculate the tension redistribution. A central unit cell is cropped and used for the band structure calculations (**B**) Corresponding mesh of the unit cell.

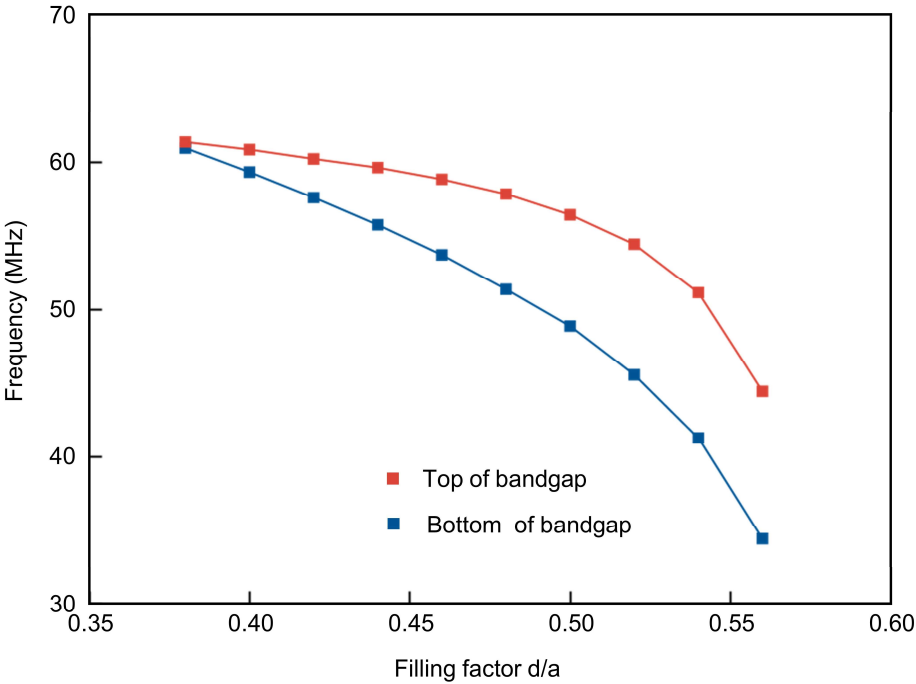

**Figure S5. Phononic Bandgap vs. filling factor.** Frequencies of the top and of the bottom of the bandgap vs. the filling factor d/a for a = 1 µm and an initial tension of $T_0$ = 0.01 N/m. Tension redistribution is accounted for.

For the calculations in the **finite model** we use a membrane model from the solid mechanics module with two study steps (Figure S6). In the first step, we again let the system relax after adding uniform built in tension. The resulting tension distribution is shown in Figure S6 C and D. We then calculate the first 1500-2500 resonances of the system in an eigenfrequency study step. The mode shapes and frequencies are exported for further analysis in a python script (see section III). Also here it is important for the mesh to represent the symmetry of the modelled geometry – compare Figure 6B.

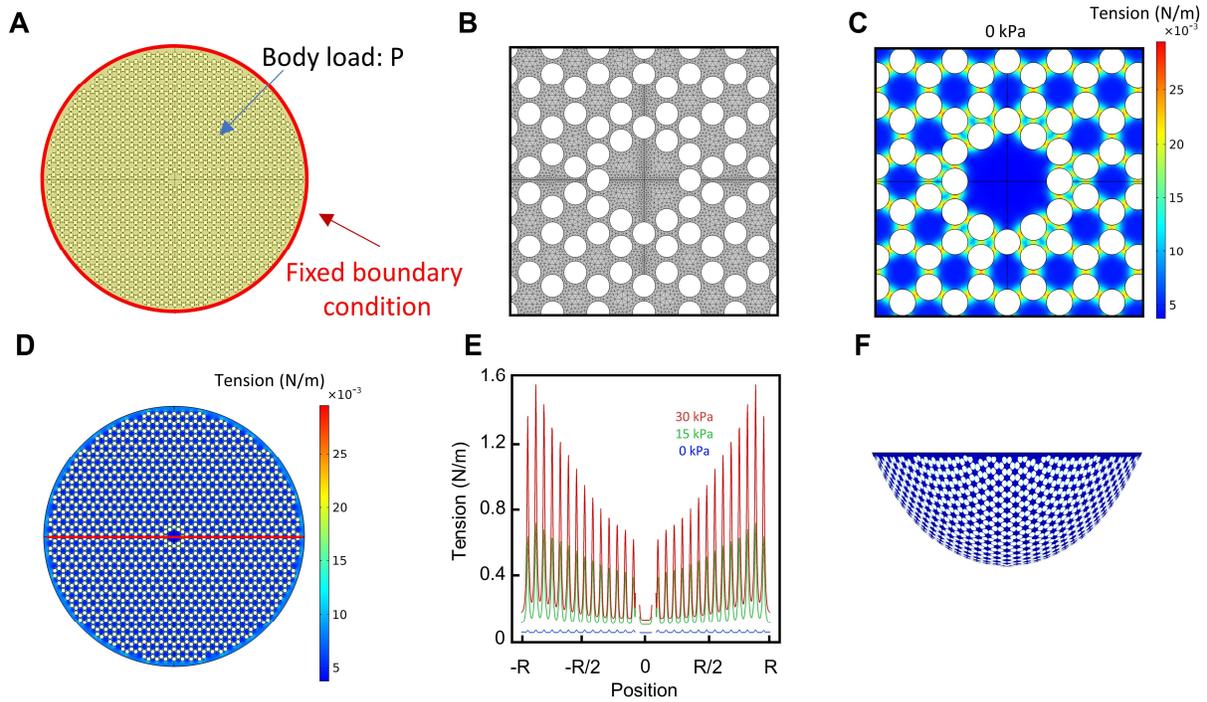

**Figure S6. Finite model PnC simulations.** (**A**) Circular PnC of 30.6 μm size. The electrostatic pressure is applied as a body load (yellow). (**B**) Symmetric mesh. (**C**) Tension after the redistribution step around the central defect region and in the entire device (**D**). (**E**) Tension along a line cut (marked red in (**D**)) for different pressures. With larger pressures, the distribution becomes increasingly non-uniform. (**F**) Deformation of the finite system under large pressures (100 kPa) to illustrate the perturbation of the phononic pattern.

III.   Mode shape analysis

We first export the mode shape for each mode obtained from our Comsol simulations and interpolate it onto a square grid with 1000 nodes. Next, we carry out a fast Fourier transform (FFT) on such an array to obtain the reciprocal space representation of each mode. Most modes, except for the ones within the bandgap, have a well-defined momentum along each specific direction. To determine the momentum content of each mode, we take a cut of each mode in the momentum space and find a peak along each particular direction. To aid in this procedure and reduce noise, we average over 10 neighboring modes. Knowing the momentum, we finally export the dispersion relation along the direction of interest.

IV.    **Raman spectroscopy analysis**

In the main paper, we use initial tension $T_0$ as a device parameter to avoid confusion. For Raman data however it is more common characterize graphene in terms strain $\epsilon_0$, which is directly linked to tension value via the 2D-Youngs-modulus $T_0 = \epsilon_0 E_{2D}$. In this section, we discuss in detail the signatures of strain redistribution obtained by Raman spectroscopy of the graphene phononic crystal presented in the main paper. Figure S7A shows a Raman map of the integrated 2D-mode intensity of the suspended membrane, see Figure 1D of the main paper. The holes forming the phononic crystal are clearly marked by a local decrease in 2D-mode intensity. We show a representative Raman spectrum from the centre of the phononic crystal in Figure 7B, marked by (#) in Figure S7A. The experimentally observed intensity ratio between the Raman G and D peaks, I(2D)/I(G)>1, proves that the phononic crystal is made from a single layer of graphene. We used SEM imaging to avoid measuring the devices with bilayer graphene areas (Fig. S3A). The appearance of the D, D' and D+D' mode indicate the presence of defects, which arise mainly from repeated electron beam imaging of the graphene membrane.

To demonstrate the onset of strain relaxation, we focus on a horizontal line cut (along x within out laboratory frame) across the phononic crystal at y = 2.3 µm as indicated by the dashed line in Figure S7A. Figure S7C shows the corresponding integrated 2D-mode intensity (squares) and position (triangles) as a function of x, where the origin (0,0) was set at the centre of the membrane. We observe four equidistant drops in intensity, indicated by arrows, which corresponds to the narrow graphene stripes between the holes (compare to Figure S7A). The drop in intensity occurs because at the strips, the laser spot overlaps with the holes in the graphene membranes such that less material is probed compared to regions further away from the holes. For the two narrow graphene stripes closest to 0, we find that the 2D-mode position drops by 3-5 cm$^{-1}$ compared to the central region of the graphene membrane. As hydrostatic strain lowers the energy of phonons probed by Raman spectroscopy,[2,3] our observation suggests that the narrow stripes carry a higher strain then neighbouring parts of the graphene phononic crystal, which is in qualitative agreement with the strain redistribution introduced in the main paper. A similar but less pronounced behaviour occurs for the two additional narrow stripes closer to the edge of the suspended graphene membrane at $x \sim \pm 3\mu m$, where the local decrease in 2D-mode position (increase in strain) is overlaid with a general increase of the 2D-mode position (decrease in strain) towards the edge of the suspended part of the graphene membrane at $x \sim \pm 4\mu m$. The overall lower position 2D-mode in the centre of the membrane could be due to laser heating. The pattering reduces the thermal conductance of the system and thus even at small powers (0.5 mW) heating can occur.

Next, we quantify the hydrostatic strain in our phononic crystal, which is presented in Figure 1E of the main paper. Hydrostatic strain $\varepsilon_h$ in graphene leads to a shift $\Delta\omega_{2D}$ of the 2D-mode position $\omega_{2D}$ following the relation.[2,3]

$$\Delta\omega_{2D} = -\varepsilon_h \, \gamma_{2D} \, \omega_{2D}^0 \quad (1)$$

where $\gamma_{2D} = 2.6$ is the Grüneisen parameter of the 2D-mode in graphene, and $\omega_{2D}^0$ is the intrinsic 2D-mode position without strain or doping ($\omega_{2D}^0 = 2678 \, cm^{-1}$ for $532 \, nm$ excitation) [4]. From the measured 2D-position $\omega_{2D}^{epx}$, $\Delta\omega_{2D} = \omega_{2D}^0 - \omega_{2D}^{epx}$, and equation (1) we extract the strain values from Figure S7C and show them together with the 2D-mode position in Figure 1 of the main paper.

The general trend of lower strain towards the edge of the suspended graphene phononic crystal suggests that strain relaxation is not complete across the entire structure. We attribute this behaviour to two main reasons. First, strain in suspended graphene membranes is never homogeneous, see reference membrane in Figure 1D of the main paper, and the degree of strain relaxation scales with the absolute strain values. Therefore, we do not expect homogenous strain relaxation across the entire phononic crystal. Second, strain in suspended graphene visibly varies on length scales comparable to the size of the holes in the suspended graphene membrane, see Figure 1D of the main paper, which makes strain relaxation less effective. Here we chose a PnC with rather large lattice constant $a$ such that strain variation and relaxation can be probed by Raman spectroscopy with diffraction limited spatial resolution. For phononic crystals with holes sizes and periodicities that are much smaller than the variations of initial strain in suspended graphene structures, we expect strain relaxation to be more efficient than what is observed for the phononic crystal discussed here.

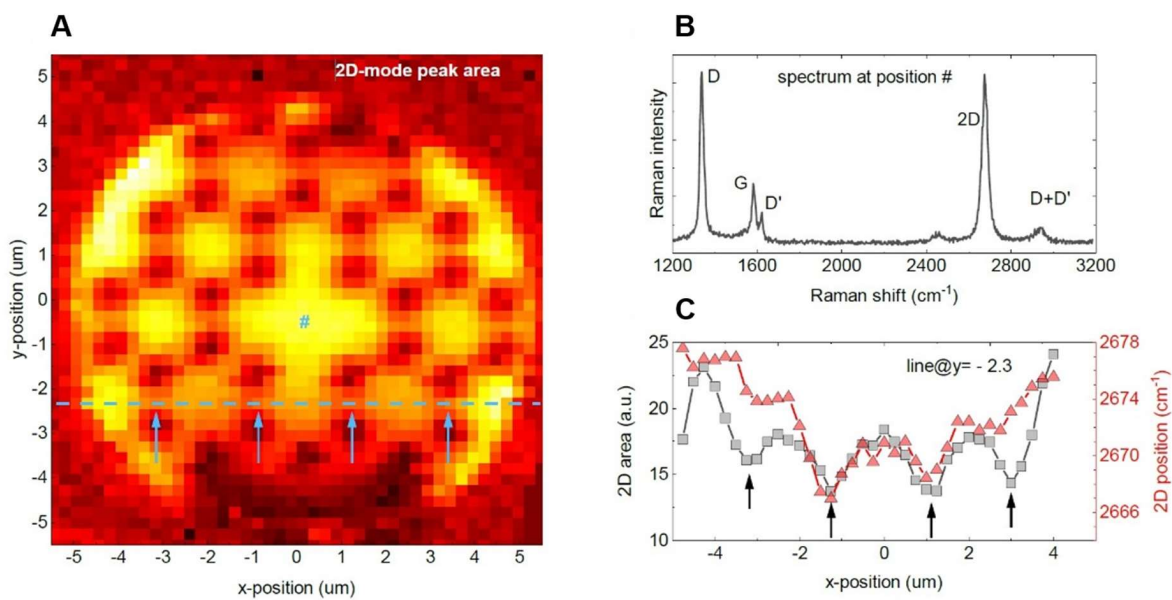

**Figure S7. Raman characterization of the graphene PnC. (A)** Raman map of the graphene 2D-Raman mode (integrated intensities. Intensity drops mark the locations of the holes in the hexagonal arrangement that forms the phononic crystal. **(B)** Representative Raman spectrum of the graphene membrane extracted at the location marked as (#) in **(A)**. The dominant Raman modes of graphene are labelled. **(C)** Integrated 2D mode intensity (area, grey squares) and 2D-mode position (red triangles) along a line cut at y = 2.3um in x-direction as indicated by the dashed line in **(A)**. Arrows in **(A)** and **(C)** mark the locations of narrow graphene bridges between the holes where strain relaxation is expected.

## V. Experimental signatures of the defect mode

We propose detecting the defect mode by interferometric detection. In this approach a laser beam reflected from the device interferes with a reference beam providing an accurate measurement of membrane's position.[5–8] We need to confirm, however, that diffraction-limited laser spot is small enough to measure signatures of a realistic defect mode. To confirm that the defect mode in the center of the PnC presented in the main text is detectable, we simulate the spatial signal read out by the interferometer by implementing a Gaussian averaged laser spot for a realistic source reflected from our structure. In Figure S8 we show this for multiple laser spot sizes. For the smallest possible spot with a FWHM of 720 nm, even small spatial features of the mode shape are detectable (Figure S8A). For a realistic spot size (FWHM of 2400 nm) including the window of a vacuum chamber and a larger working distance objective we are still able to measure the mode (Figure S8B). And finally, we take a very large spot (FWHM of 7.2 μm) and thereby probe the entire mode shape (Figure S8C). We can confirm that the mode will not be fully averaged out, as it has net displacement (in contrast to e.g. mirror symmetric modes). Overall, we confirm that we should be able to detect the motion of the defect mode for all realistic laser spot sizes.

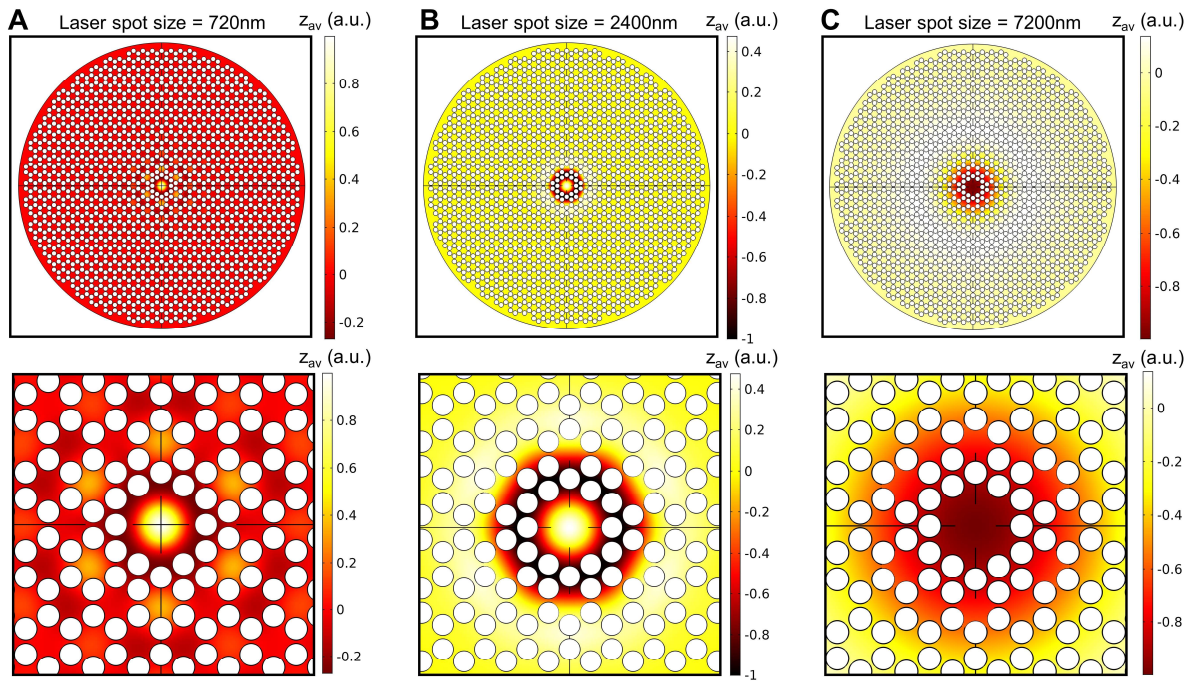

**Figure S8. Interferometric detection of the defect mode**. (**A-C**) Mode shape of the defect mode with local Gaussian averaging to simulate the displacement detection via a focused laser spot of different width (Zoom-ins are shown below). Different panels correspond to different spot sizes. Even for the largest laser spot size a net displacement is evident.

As mentioned in the main text spatial uniformity is necessary to fabricate a phononic crystal with a well-defined band structure. Monolayer graphene is sensitive so surface effects, wrinkling and fabrication residues. Using multilayer graphene would solve this problem yet will also be less responsive to the experimentally possible pressure maximum of roughly 30 kPa. To check if a PnC made from multilayer graphene would still show frequency tuning, we simulate the resonance frequency of a uniform circular membrane (initial tension 0.01 N/m) with and without applied load of 30 kPa. In Figure S9 we plot the relative frequency change under pressure vs. number of graphene layer. Even though the tunability drops quickly for thicker graphene, we still find more than 100% possible upshift for 35 layers.

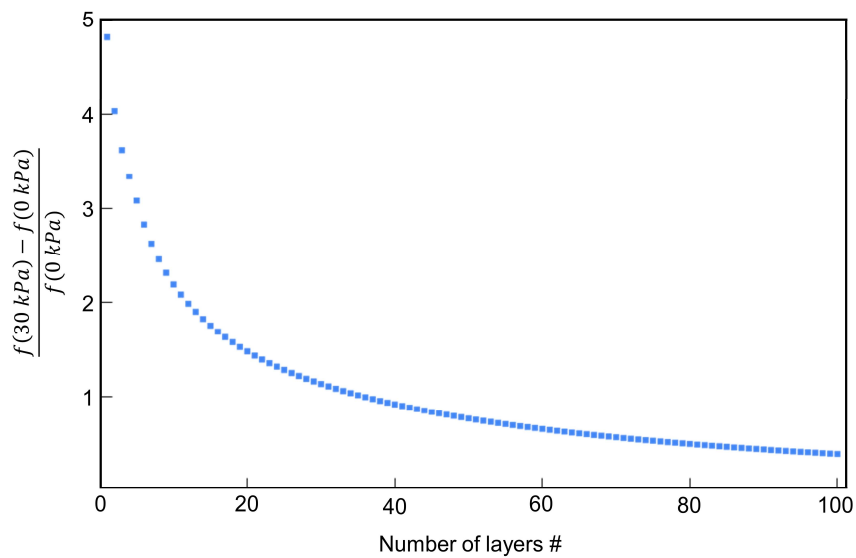

**Figure S9. Frequency tunability vs. number of layers.** Relative frequency shift of the fundamental mode of a circular multilayer graphene resonator upon applying 30 kPa of pressure vs. number of graphene layers.